\documentclass[%
 reprint,
 amsmath,amssymb,
 aps,usenames,dvipsnames
floatfix,
]{revtex4-2}

\usepackage{natbib}
\usepackage{graphicx}
\usepackage{dcolumn}
\usepackage{bm}
\usepackage{ulem}
\usepackage{hyperref}
\usepackage{cleveref}

\begin{document}

\preprint{APS/123-QED}

\title{Biaxial Strain Modulated Valence Band Engineering in III-V Digital Alloys}

\author{Sheikh Z. Ahmed}
\email{sza9wz@virginia.edu}
 \affiliation{Department of Electrical and Computer Engineering, University of Virginia, Charlottesville, Virginia 22904, USA}
\author{Yaohua Tan}%
\affiliation{Synopsys Inc, Mountain View, California 94043, USA}

\author{Jiyuan Zheng}
\affiliation{Beijing National Research Center for Information Science and Technology, Tsinghua University, 100084, Beijing, China}%

\author{Joe C. Campbell}
\affiliation{Department of Electrical and Computer Engineering, University of Virginia, Charlottesville, Virginia 22904, USA}

\author{Avik W. Ghosh}
\affiliation{Department of Electrical and Computer Engineering, University of Virginia, Charlottesville, Virginia 22904, USA}
\affiliation{Department of Physics, University of Virginia, Charlottesville, Virginia 22904, USA}

\date{\today}

\begin{abstract}
Some III-V digital alloy avalanche photodiodes exhibit low excess noise. These alloys have low hole ionization coefficients due to presence of small 'minigaps', enhanced effective mass and large separation between light-hole and split-off bands in the valence band. In this letter, an explanation for the formation of the minigaps using a tight binding picture is provided. Furthermore, we demonstrate that decreasing substrate lattice constant can increase the minigap size and mass in the transport direction. This leads to reduced quantum tunneling and phonon scattering of the holes. Finally, we illustrate the band structure modification with substrate lattice constant for other III-V digital alloys.  
\end{abstract}

\maketitle


\section{\label{sec:Introduction}Introduction}
The fields of silicon photonics, telecommunication and light imaging, detection and ranging (LIDAR) systems are undergoing unprecedented growth with the emergence of the Internet of Things, spawning a correspondingly increased demand for efficient photodetectors \cite{Tosi, CAMPBELL2008221, bertone2007avalanche, Mitra2006, Williams2017, Nada2020, Pasquinelli2020, Thomson_2016}. Avalanche Photodiode (APD) is an ideal candidate for such applications due to its intrinsic gain mechanism which enables higher sensitivity \cite{apd_recent}. However, the gain performance of an APD is associated with excess noise which arises due to the stochastic nature of the impact ionization process. The excess noise factor $F(M)$ is set by the variance in particle count $\sigma_m^2 = \langle m^2\rangle - \langle m\rangle^2$ vs the mean particle gain $\langle m\rangle = M$  through the relation $\sigma_M^2/M^2 = F(M)-1 =  (M-1)/M + k(M-1)^2/M$. The average particle strength $\langle m^2\rangle$ in turn contributes to the shot noise current fluctuation $\langle i_{shot}^{2}\rangle =2qIM^{2} F(M) \Delta f$ \cite{mcintyre1966multiplication,Teich1986,Teich1990}. Here, $q$ is the electron charge, $I$ is the total photo plus dark current, $M$ the average multiplication gain and $\Delta f$ the bandwidth. The excess noise can be minimized by reducing the ratio $k$ of hole ionization coefficient $\beta$ to electron ionization coefficient $\alpha$ for electron injected APDs. For hole injected APDs the ratio is reversed. Primarily, we can reduce the excess noise in three possible ways - choosing semiconductor materials with favorable impact ionization coefficients, adjusting the multiplication region width to utilize the non-local aspect of impact ionization, and designing heterojunctions in order to engineer the impact ionization process \cite{apd_recent}. 

III-V digital alloy APDs with very low excess noise and high gain-bandwidth product operating in the short-infrared wavelength spectrum have recently been reported \cite{InAlAs_expt,AlInAsSb_expt,AlAsSb_expt}. Digital alloys are short-period superlattices that include alternately stacked binary compounds in a periodic fashion. The low $k$ in these few digital alloys can be ascribed to multiple factors -  the generation of 'minigaps' in the material valence band, a corresponding enhanced valence band effective mass and finally a large separation between the light-hole and split-off bands \cite{AlInAsSb_MC, InAlAs_MC,InAlAs_MC2}. These properties prevent holes from gaining energy. keeping them localized near the valence band edge. However, in these electron injected APDs, electrons in the conduction band can easily move to higher energies, bypassing conduction band minigaps, in order to impact ionize due to their low effective mass. 

Minigaps are seen to arise naturally in the first-principles unfolded bandstructures calculated for the superlattice stack. However, their chemical origin is not well understood and require an in-depth analysis. While the presence of minigaps is not a necessary condition for high photogain with low excess noise, it may well prove to be a sufficient condition in many cases. It is thus useful to identify ways to engineer such minigaps deterministically with various design knobs, such as alloying and strain.

In this paper, we use a simple $sp^3$ tight-binding model to illustrate how strain alters the bonding chemistry in APD digital alloy materials and plays a crucial role in the formation of minigaps. We then employ a more elaborate Environment-Dependent Tight Binding (EDTB) model \cite{TanETB,TanSi} with band unfolding techniques \cite{tan_unfolding,boykin_unfolding1,boykin_unfolding2} to investigate the role of strain in the formation and modulation of these digital alloy minigaps.  Furthermore, we study the relationship between biaxial strain and minigap size and their overall impact on carrier transport. This study provides a convenient design principle towards efficient photodetectors, and for overall tunability of electron wavefunction in digital superlattices.

\section{Simulation Method}
A digital alloy bandstructure requires particular attention to two multiscale attributes - short range atomistic modifications at the hetero-interfaces, and long-range band modulation by the superlattice potential.
Conventional tight binding models are typically not suitable for handling the material chemistry at interfaces and surfaces, as they are calibrated to the higher order symmetry of the bulk crystallographic point group \cite{TanSi}.
One alternative is to use non-orthogonal tight binding approaches such as Extended H\"uckel Theory with explicit atomic basis sets, generating parameters that are transferrable between diverse environments such as bulk vs severely strained and reconstructed surfaces \cite{eht1}, \cite{eht2}.
As an alternate, the tight binding parameters of our EDTB model are explicitly environment-dependent, calibrated to state-of-the-art Hybrid Density Functional Theory (HSE06) \cite{heyd2003hybrid} band-structure as well as their orbital resolved wavefunctions. As a result, the model is able to accurately capture changes due to effects like strain and interface reconstruction by tracking changes in the atomic coordinates, bond lengths and angles. 

While atomistic details are captured by our EDTB, we still have to deal with the large unit cells of our digital alloys, which generate an aggressively scaled Brillouin zone and a very complicated bandstructure due to the proliferation of a large number of zone-folded bands \cite{AhmedTFET}. The sphagetti-like bandstructure is simplified by employing a band-unfolding technique \cite{tan_unfolding,boykin_unfolding1,boykin_unfolding2}. In this method, the supercell bands are projected onto the Brillouin zone of a primitive cell, with weights set by decomposing individual eigenfunctions into several Bloch wavefunctions with separate wave vectors of the primitive cell Brillouin zone. 

When it comes to simulating carrier transport, it is worth keeping in mind that there are two primary mechanisms by which carriers can bypass the minigaps, namely, quantum tunneling and optical phonon scattering. The impact of tunneling through a minigap is readily captured by computing the ballistic transmission in 3-D using the Non-Equilibrium's Green's Function (NEGF) formalism. The 3-D Hamiltonian is broken into nearest-neighbor blocks along the transport direction, including the applied electric field, while Fourier transforming in the transverse directions to capture in-plane structural periodicity. At the two ends of the APD, we consider extensions of the material at constant potential, generating a semi-periodic array of transport blocks with onsite matrix $\alpha_{i\vec{k}\perp}$ ($i=1,2$ for the  bias-separated contacts at the two ends) and hopping matrix $\beta_{\vec{k}\perp}$. Using recursion, we find the two contact surface Green's functions $g$ and then the self-energy matrices $\Sigma$, whose anti-Hermitian parts give us the broadening matrices $\Gamma$ related to the electron escape rates into the contacts. We can then calculate the retarded Green's Function $G$ and the quantum mechanical transmission $T$ using the Fisher-Lee formula \cite{ghosh}
\begin{eqnarray}
g^{-1}_{1\vec{k}\perp} &=& \alpha_{1\vec{k}\perp} -  \beta^\dagger_{\vec{k}\perp}g_{1\vec{k}\perp}\beta_{\vec{k}\perp}, ~~~\Sigma_{1\vec{k}\perp} = \beta^\dagger_{\vec{k}\perp}g_{1\vec{k}\perp}\beta_{\vec{k}\perp} \nonumber\\   
g^{-1}_{2\vec{k}\perp} &=& \alpha_{2\vec{k}\perp} -  \beta_{\vec{k}\perp}g_{1\vec{k}\perp}\beta^\dagger_{\vec{k}\perp},~~~\Sigma_{2\vec{k}\perp} = \beta_{\vec{k}\perp}g_{2\vec{k}\perp}\beta^\dagger_{\vec{k}\perp} \nonumber\\  
\Gamma_{i\vec{k}\perp} &=& i\left(\Sigma_{i\vec{k}\perp} - \Sigma^\dagger_{i\vec{k}\perp}\right)\nonumber\\
G_{\vec{k}\perp}(E) &=& \left[(E+i0^+)I - H_{\vec{k}\perp} - \Sigma_{1\vec{k}\perp} -\Sigma_{2\vec{k}\perp} \right]^{-1}\nonumber\\
T &=& \sum_{\vec{k}\perp}Tr\left(\Gamma_{1\vec{k}\perp}G_{\vec{k}\perp}\Gamma_{2\vec{k}\perp}G^\dagger_{\vec{k}\perp}\right)
\end{eqnarray}
The NEGF formalism uses a Hamiltonian from the EDTB model with elements being represented in the basis set of transverse momenta $\vec{k}_\perp$ perpendicular plane to transport direction. This is accomplished by starting with a 3-D bulk $E-\vec{k}$ and inverse transforming along the transport direction to naturally generate the tridiagonal $\alpha_{\vec{k}\perp}$, $\beta_{\vec{k}\perp}$ blocks \cite{stovneng1993multiband}. 

Using this model, we compute the energy dependent ballistic transmission to see how minigap size impacts the quantum tunneling process. Including phonon scattering in NEGF typically requires generalizing from Fisher-Lee to the Meir-Wingreen incoherent transport formulation with an added phonon self-energy obtained within a self-consistent Born approximation \cite{ghosh}. Instead,
we study the effect of phonon scattering in the digital alloys using a multi-band Boltzmann transport model that focuses on classical transport with the quantum physics hidden in the band parameters. The model outputs the energy resolved carrier occupation probability, which we calculate to explore the effect of minigaps on phonon scattering.

\section{Formulation of Theory}

\begin{figure*}[t!]
\includegraphics[width=0.75\textwidth]{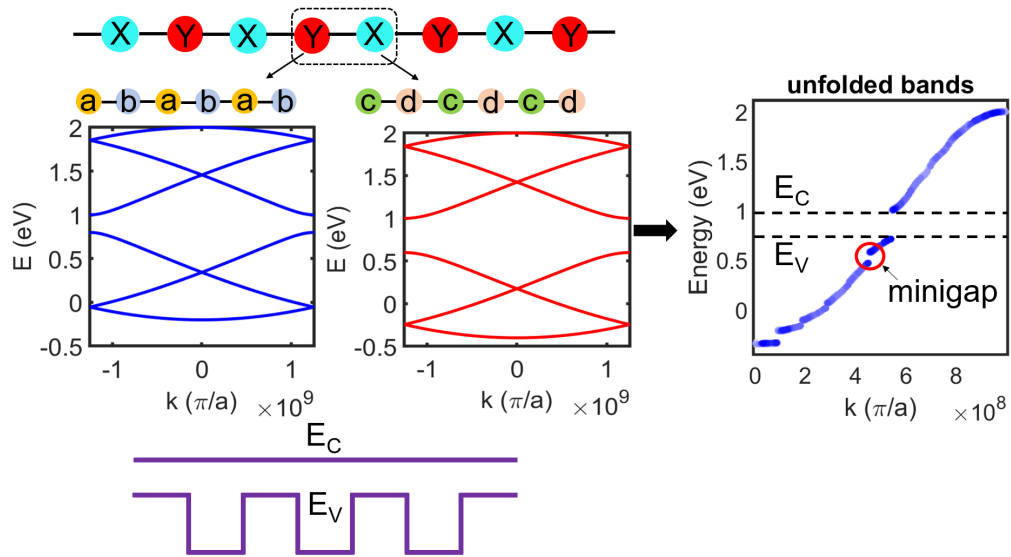}
\caption{\label{toy_model} In a toy bandstructure model, we consider an unit cell (dotted box) consisting two arbitrary binary materials. By adjusting onsite energies and hopping parameters a material system with zero conduction band offset and large valence band offset is created. The large valence band offset results in minigaps in the valence band (within red circle) as shown in the unfolded band structure.
}
\end{figure*}

Previous studies of digital alloys like InAlAs, AlInAsSb and AlAsSb \cite{InAlAs_MC,InAlAs_MC2,AlInAsSb_MC,InAlAs_expt} have demonstrated that valence band minigaps present in the material bandstructure play a part in reducing excess noise by limiting hole carrier transport. However, the role of minigap is firmly established for one material combination, InAlAs. For the other materials either a systematic experimental comparison between digital and random alloy superlattices does not exist, or when it does, the random shows low noise as well and is attributed to an energy separated split-off band \cite{sahmed}. Nonetheless, a deterministic creation of a strong minigap can significantly aid APD gain by suppressing one carrier type. 
In this section, we explore the formation of these gaps using a one dimensional simple 'toy' model.

We consider an arbitrary alloy consisting of two materials $X$ and $Y$ stacked alternately like a digital alloy, as shown in Fig.~\ref{toy_model}. Each of these materials is essentially a dimer consisting of a set of two atoms. For $X$, the component atoms are $a$ and $b$, and for $Y$ they are $c$ and $d$. The resulting Hamiltonian of the unit cell for this material then looks like: 

\begin{equation}
H=\left(\begin{array}{cccccc}
\alpha_X & -\beta_X & & & & -\gamma_{YX1}\\
-\beta^{\dagger}_{X} & \alpha_X & -\beta_X &\\
& -\beta^{\dagger}_{X} & \alpha_X & -\gamma_{XY1}\\
& & -\gamma_{XY2} & \alpha_Y & -\beta_Y\\
 &  &  & -\beta^{\dagger}_{Y} & \alpha_Y & -\beta_Y \\
-\gamma_{YX2} &  &  &  & -\beta^{\dagger}_{Y} & \alpha_Y\\
\end{array}\right)
\end{equation}
where, 
\begin{equation}
\alpha_X=\left(\begin{array}{cc}
E_X & -t_1\\
-t_1 & E_X
\end{array}\right),\beta_X=\left(\begin{array}{cc}
0 & 0\\
t_2 & 0
\end{array}\right)
\end{equation}

\begin{equation}
\alpha_Y=\left(\begin{array}{cc}
E_Y & -t_3\\
-t_3 & E_Y
\end{array}\right),\beta_Y=\left(\begin{array}{cc}
0 & 0\\
t_4 & 0
\end{array}\right)
\end{equation}

\begin{equation}
\gamma_{XY1}=\left(\begin{array}{cc}
0 & 0\\
t_5 & 0\\
\end{array}\right),\gamma_{XY2}=\left(\begin{array}{cc}
0 & t_6\\
0 & 0
\end{array}\right)
\end{equation}

\begin{equation}
\gamma_{YX1}=\left(\begin{array}{cc}
0 & t_7\\
0 & 0\\
\end{array}\right),\gamma_{YX2}=\left(\begin{array}{cc}
0 & 0\\
t_8 & 0
\end{array}\right)
\end{equation}
For each material, we consider the onsite energies, $E_{X,Y}$ to be constant, while the hopping parameters $t_{1,2,3,4}$ between the dimer elements vary. $t_{5,6,7,8}$ represent the coupling between material $X$ and material $Y$. Here, we set $E_X=0.9$, $E_Y=0.8$, $t_1=0.6$, $t_2=0.5$, $t_3=0.7$, $t_4=0.5$, $t_5=-0.4$, $t_6=-0.6$, $t_7=-0.4$ and $t_8=-0.6$ in eV. These parameter values are chosen such that there is a large valence band offset between $X$ and $Y$ but the conduction band offset is zero, as depicted in Fig.~\ref{toy_model}.  The resulting unfolded bandstructure is shown on the right side of the figure. We observe that a clear minigap forms in the valence band (highlighted with a red circle), while correspondingly large minigaps do not arise in the conduction band. This simple example illustrates that sizeable minigaps can be engineered selectively in one band by creating large onsite energy variations in the frontier atomic orbitals that generate that band. We will now explore how such large offsets can be deliberately engineered in the III-V digital alloys using strain.

\section{Results And Discussion}
In bulk heterojunctions, band discontinuities form at the interface owing to the alignment of Fermi levels of the constituent components, resulting in band offsets. The band offset sizes can be manipulated if the position of band edges can be altered \cite{peter2010fundamentals}. This is achieved by means of hydrostratic pressure \cite{Wei1999}, applying biaxial strain \cite{Pollak1968,Walle1986,Walle1989,Wei1998,Kent2002} and alloying \cite{bir1974symmetry,Mailhiot1989}. In digital alloys it is biaxial strain that results in the opening of the minigaps, as we will describe next. 

\begin{figure}[t]
\includegraphics[width=0.42\textwidth]{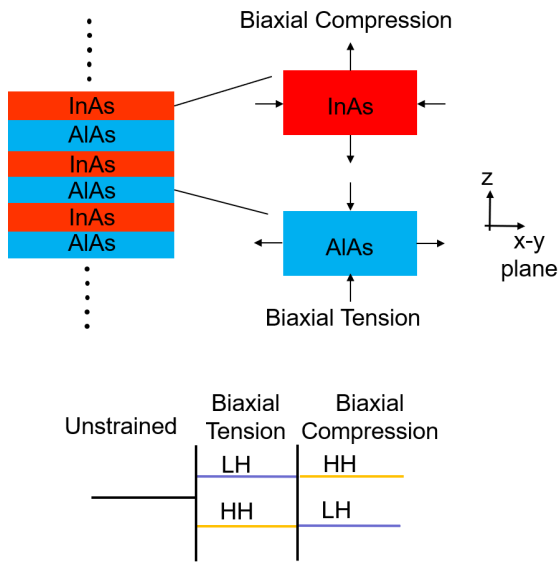}
\caption{\label{minigap_explanation} 6-monolayer InAlAs consists of InAs and AlAs grown on InP substrate. Thus, InAs experiences biaxial compression and AlAs experiences biaxial tension. Biaxial strain results in splitting of the HH and LH bands. Since InAs and AlAs experience opposite kinds of strain, their bands move in opposite direction. This results in opening of minigaps in InAlAs.   
}
\end{figure}

It is well known that biaxial strain in semiconductors removes the degeneracy of the valence bands and results in the splitting of the heavy-hole (HH) and light-hole (LH) bands \cite{Mailhiot1989,Nishida2007}. Let us consider the case of InAlAs to understand how minigaps form. In Fig.~\ref{minigap_explanation}, it is shown that InAlAs consists of InAs and AlAs stacked alternately. The alloy is grown on an InP substrate having a lattice constant $5.87${\AA} \cite{TanETB}. Compared to InP, the lattice constant of InAs at $6.06${\AA} is greater while that of AlAs at $5.66${\AA} is smaller. As a result, AlAs experiences biaxial tension in the $x-y$ plane, while InAs undergoes biaxial compression. In the (001) $z$ -direction, InAs undergoes expansion and AlAs undergoes compression. As we will see shortly, biaxial tension results in LH bands moving up and HH moving down in energy, as depicted at the bottom of Fig.~\ref{minigap_explanation}. The opposite happens for biaxial compression. As the bands in the alternately strained layers move in opposite directions, the band offset increases, resulting in the formation of the minigaps. 

\begin{figure}[t]
\includegraphics[width=0.42\textwidth]{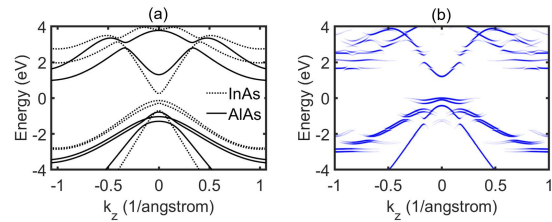}
\caption{\label{InAlAs_bandstructure} (a) Bandstructure of InAs and AlAs grown on InP substrate (b) unfolded bandstructure of 6ML InAlAs with InP as the substrate. The bandstructures above are calculated using the sp$^3$s$^*$d$^5$ EDTB model. 
}
\end{figure}

Fig.~\ref{InAlAs_bandstructure}(a), shows the bandstructure of the strained InAs and AlAs (grown on InP substrate) computed with the $sp^{3}s^{*}d^{5}$ EDTB model. We observe a large valence band offset at the $\Gamma$ point between the strained AlAs and InAs. The unfolded bandstructure of a 6-monolayer InAlAs showing the resulting valence band minigaps, computed with the EDTB model, is depicted in Fig.~\ref{InAlAs_bandstructure}(b). The unit cell of the InAlAs DA considered consists of 3ML AlAs and 3ML InAs. In order to comprehend the movement of these bands, a closer look at the orbital chemistry is required.

\begin{figure}[b]
\includegraphics[width=0.42\textwidth]{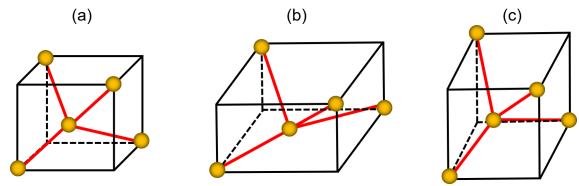}
\caption{\label{biaxial_types} (a) Unstrained zinc-blende crystal, (b) under biaxial tension in the $x-y$ plane, and (c) under biaxial compression in the $x-y$ plane.
}
\end{figure}

In a bulk zinc blende semiconductor, each atom is tetrahedrally bonded to four neighboring atoms. The bonds connecting these atoms point toward the $\langle 111 \rangle $ directions of the cube that bounds around the tetrahedron. Every bond consists of 25\% contribution each from the $s$, $p_x$, $p_y$ and $p_z$ orbitals \cite{Nishida2007}. Fig.~\ref{biaxial_types}(a) shows the chemical bonds in the unit cell of an unstrained zinc-blende crystal. The bonds
have cubic point group symmetry, so the valence bands are degenerate at the $\Gamma$ point.  However, under biaxial tension (uniaxial compression along $z$) all the bonds are equally rotated towards the $x-y$ plane (Fig.~\ref{biaxial_types}b), while under biaxial compression they move away from the $x-y$ plane (Fig.~\ref{biaxial_types}c). Near the valence band edge, bonding states arising from the overlap of the directional $p$ orbitals mainly contribute to the formation of the bands there. The spherical $s$ orbitals contribute to the conduction band edge states. Considering only contributions from the $p$ orbitals and projecting one of the tetrahedral bonds along a principal direction ($\langle 100 \rangle $, $\langle 110 \rangle $ or $\langle 111 \rangle $), the out-of-plane orbital forms the LH states, i.e., $p_z$ orbital along $(001)$ or $z$ direction. Then, HH states are formed by the in-plane orbitals, for instance, $p_x$ and $p_y$ orbitals if we are looking from the $z-$direction. We can then explain the effect of strain on these $p$ orbitals using a simple $sp^3$ tight binding model. The ignored virtual s$^*$ and d orbitals end up being important quantitatively, the former for indirect band-gap semiconductors like Si, the latter to nail down its transverse effective masses. However, they have less qualitative relevance to direct bandgap III-V materials. We use the full sp$^3$s$^*$d$^5$ set for our numerical evaluations, but a simplified sp$^3$ for the current qualitative arguments.

Chadi and Cohen \cite{chadi1975} and Harrison \cite{harrison2012electronic} used $sp^3$ tight-binding model to investigate the electronic band structure of various diamond and zinc-blende crystals. In  the model, the valence band orbitals form the the conduction and valence bands. Each atom in the primitive cell contributes an $s$, $p_x$, $p_y$ and $p_z$ orbital. The resulting Hamiltonian is an $8\times 8$ matrix without inclusion of spin-orbit coupling. At the $\Gamma$ point, the $sp^3$ Hamiltonian can be simplified to:

\begin{equation}
H=\left(\begin{array}{cccccccc}
E_{SC} & V_{SS} & 0 & 0 & 0 & 0 & 0 & 0\\
V_{SS} & E_{SA} & 0 & 0 & 0 & 0 & 0 & 0\\
0 & 0 & E_{PC} & V_{XX} & 0 & 0 & 0 & 0\\
0 & 0 & V_{XX} & E_{PA} & 0 & 0 & 0 & 0\\
0 & 0 & 0 & 0 & E_{PC} & V_{YY} & 0 & 0\\
0 & 0  & 0 & 0 & V_{YY} & E_{PA} & 0 & 0\\
0 & 0 & 0 & 0 & 0 & 0 & E_{PC} & V_{ZZ}\\
0 & 0 & 0 & 0 & 0 & 0 & V_{ZZ} & E_{PA}\\
\end{array}\right)
\end{equation}

\begin{figure}[t]
\includegraphics[width=0.42\textwidth]{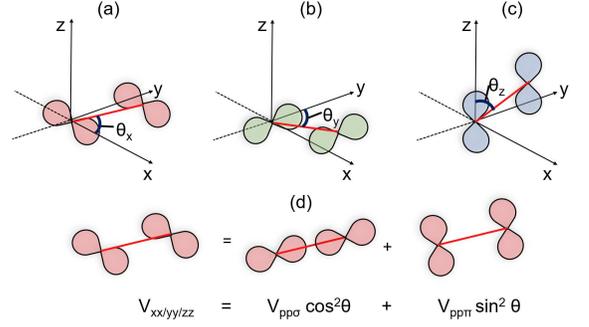}
\caption{\label{orbital_physics} Orbital overlap and azimuthal $\theta$ angle for (a) $p_x$, (b) $p_y$ and (c) $p_z$ orbitals. In (d) the $\sigma$ and $\pi$ components of the bond are shown.
}
\end{figure}

This Hamiltonian can be simplified into four $2\times 2$ matrices representing the overlap between the constituent two atoms of the four different orbitals considered. The eigenstates at the valence band edge can be computed from the Hamiltonians of the $p_x$, $p_y$ and $p_z$ orbitals:

\begin{equation}
H_1=\left(\begin{array}{cc}
E_{PC} & V_{XX}\\
V_{XX} & E_{PA}\\
\end{array}\right) H_2=\left(\begin{array}{cc}
E_{PC} & V_{YY}\\
V_{YY} & E_{PA}\\
\end{array}\right)\\ \nonumber
\end{equation}

\begin{equation}
H_3=\left(\begin{array}{cc}
E_{PC} & V_{XX}\\
V_{XX} & E_{PA}\\ 
\end{array}\right)
\end{equation}
Here, $E_{A,C}$ represent the on-site energy of the anion and cation respectively, and $V_{ii}$ is the interaction constant representing the orbital overlap. The valence band states at the $\Gamma$ point can be computed by diagonalizing these matrices to get:

\begin{eqnarray}\label{eigenstates}
    E_1 &=& \frac{E_{PC}+E_{PA}}{2}-\sqrt{\left(\frac{E_{PC}-E_{PA}}{2}\right)^2 + V_{XX}^2}\\ \nonumber
     E_2 &=& \frac{E_{PC}+E_{PA}}{2}-\sqrt{\left(\frac{E_{PC}-E_{PA}}{2}\right)^2 + V_{YY}^2}\\ \nonumber
      E_3 &=& \frac{E_{PC}+E_{PA}}{2}-\sqrt{\left(\frac{E_{PC}-E_{PA}}{2}\right)^2 + V_{ZZ}^2}\\  \nonumber
\end{eqnarray}
For an unstrained system, $V_{XX}=V_{YY}=V_{ZZ}$, which results in degenerate bands. This is consistent with the observation that bulk semiconductors are symmetric along all the cubic axes. A pictorial view of the $p_x$, $p_y$ and $p_z$ orbital overlaps is shown in Fig.~\ref{orbital_physics}(a), (b) and (c). Each $p$ orbital bond consists of head-on ($\sigma$) and side-on ($\pi$) couplings, as shown in Fig.~\ref{orbital_physics}(d). The interaction constant $V_{ii}$ is written in terms of contributions from these bonds. In the figure, $\theta$ represents the azimuthal angle between the bond and relevant  axis for the constant we are considering, i.e., $x$-axis for $V_{XX}$. These interaction constants can then be written in terms of the directional consines $(l,m,n)$ \cite{SlaterKoster}:
\begin{eqnarray}    \label{int_const}
    V_{XX} &=& l^2 V_{pp\sigma}+(1-l^2) V_{pp\pi}\\ \nonumber
    V_{YY} &=& m^2 V_{pp\sigma}+(1-m^2) V_{pp\pi}\\ \nonumber
    V_{ZZ} &=& n^2 V_{pp\sigma}+(1-n^2) V_{pp\pi}\\ \nonumber
\end{eqnarray}
For an unstrained system, $(l,m,n)=(1,1,1)/\sqrt{3}$ resulting in $V_{XX}=V_{YY}=V_{ZZ}$ and hence degenerate energy levels at the valence band edge. 

The strain tensor of a system can be broken down into three components- a hydrostatic strain and two kinds of shear strain \cite{peter2010fundamentals}. The hydrostatic strain results in overall shifting of the energy bands as the crystal symmetry is not broken. However, biaxial shear strain results in the breaking of crystal symmetry, lifting band degeneracy at the $\Gamma$ point and resulting in band warping as well. Under biaxial strain in the x-y plane, the traceless shear strain tensor can be written as
\begin{equation}
\frac{1}{3}\left(\begin{array}{ccc}
e_{xx}-e_{zz} & 0 & 0\\
0 & e_{xx}-e_{zz} & 0\\
0 & 0 & -2 \left(e_{xx}-e_{zz}\right)\\ 
\end{array}\right)    
\end{equation}
where, $e_{xx}=a_{||}/a_{i}-1$ and $e_{zz}=-D_{001} e_{xx}$. Here, $a_{||}$ and $a_{i}$ represent the substrate and epilayer lattice constants, respectively. Also, the Poisson's ratio $D=2C_{12}/C_{11}$ where $C_{11}$ and $C_{12}$ are elastic constants \cite{Walle1989}. Considering $\epsilon =e_{xx}-e_{zz}$ the directional cosines change to $(l,m,n)=(1+\epsilon,1+\epsilon,1-2\epsilon)/\sqrt{3}$. As a result, $V_{XX}=V_{YY}$, but these are not equal to $V_{ZZ}$.  Using Eq.~\ref{int_const} it is then possible to show the effect of biaxial strain on the bandstructure. 

\begin{figure}[b]
\includegraphics[width=0.42\textwidth]{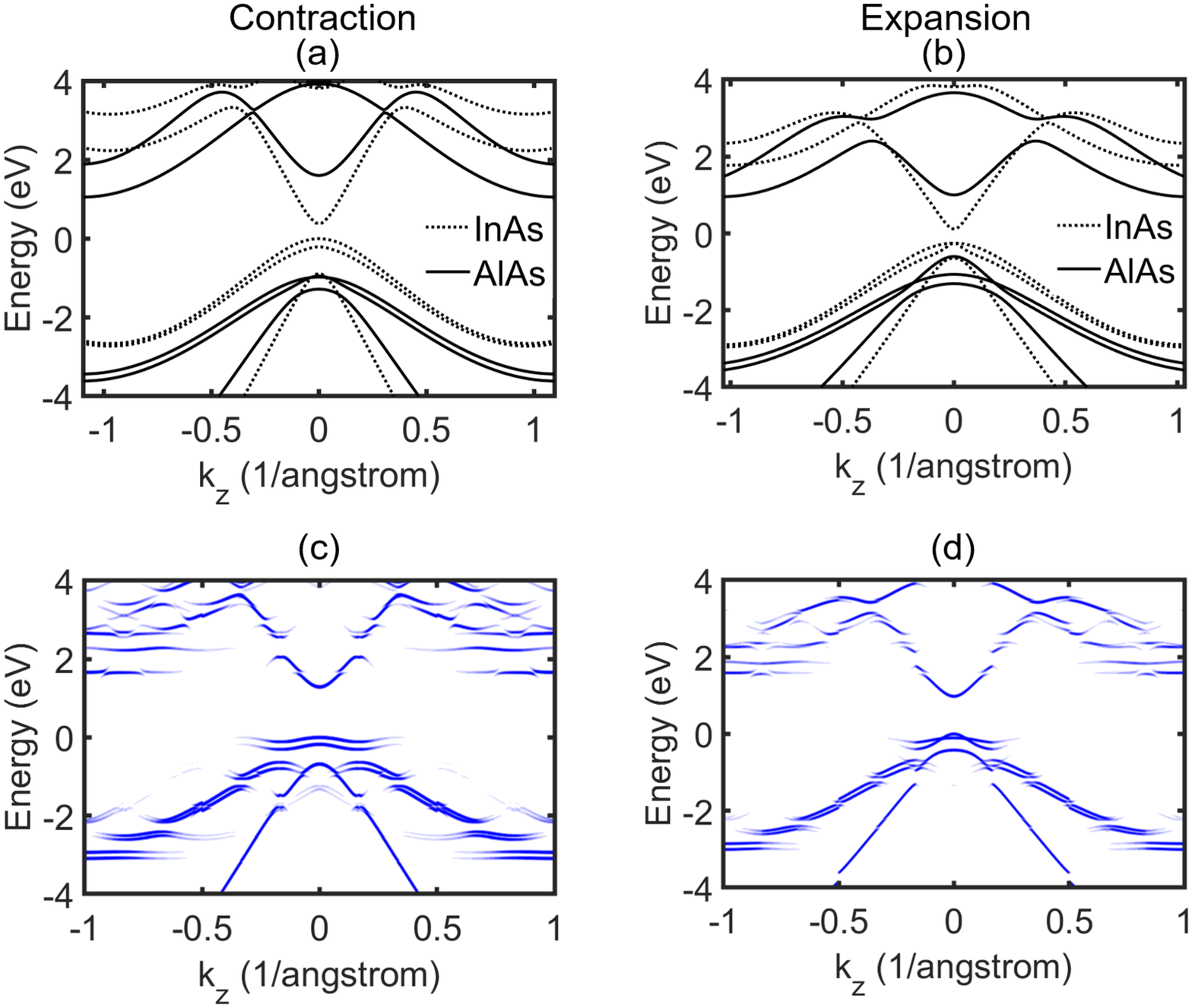}
\caption{\label{InAlAs_biaxial_stress} Bandstructure of strained InAs and AlAs for (a) ``contraction"-where substrate lattice constant is $3\%$ less than InP lattice constant and (b) ``expansion"- substrate lattice constant is $3\%$ more than InP lattice constant. The unfolded bandstructure of 6ML InAlAs under contraction and expansion is shown in (c) and (d), respectively.
}
\end{figure}

Under biaxial tension, as the bond rotates towards the $x-y$ plane, the overlap between the $p_x$/$p_y$ orbitals of the two atoms increases while overlap of the $p_z$ orbitals decreases. The azimuthal angles $\theta_x$ and $\theta_y$ decrease while $\theta_z$ increases. One can think of the $p_x$/$p_y$ orbitals of the two atoms becoming more head-on while $p_z$ orbitals becoming more parallel. This increases the contribution of the $\sigma$ components of the $p_x$/$p_y$ orbitals and weakens for the $p_z$ orbital. On the contrary, the contribution of the $\pi$ bond of the $p_z$ orbital overlap increases but diminishes for the $p_x$/$p_y$ orbitals. As a result, $V_{XX}$, $V_{YY}$ will increase while $V_{ZZ}$ will decrease, as can be inferred by placing the values of the directional cosines in Eq.~\ref{int_const}. Using Eq.~\ref{eigenstates} we can then see that the HH states go down while the LH states go up under biaxial tension. The situation is reversed under biaxial compression. The bond rotates away from the $x-y$ plane increasing $\theta_x$/$\theta_y$ and reducing $\theta_z$. This in turn leads to lower $V_{XX}$/$V_{YY}$ and higher $V_{ZZ}$. As a result, HH bands rise in energy while LH states go down. This simplified picture  explains the movement of the bands in the InAlAs digital alloy, and subsequently the essential physics of the minigap formation in the $sp^3$ basis. 

Our EDTB model, used to calculate the digital alloy bandstructures, incorporates more intricate details of higher orbitals to capture all the relevant chemistry and accurately compute material bandstructure. However, the primary underlying physics for the minigap formation, described using the $sp^3$ tight-binding model, remains the same.

\begin{figure}[t]
\includegraphics[width=0.42\textwidth]{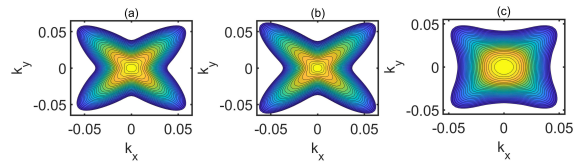}
\caption{\label{contour_xy} 2D energy contour in the $x-y$ plane of the top band of InAlAs for (a) regular (b) contraction and (c) expansion.The energy range for the contour is from $0.025eV$ to $0.5eV$ below the valence band edge.}
\end{figure}

\begin{figure}[b]
\includegraphics[width=0.42\textwidth]{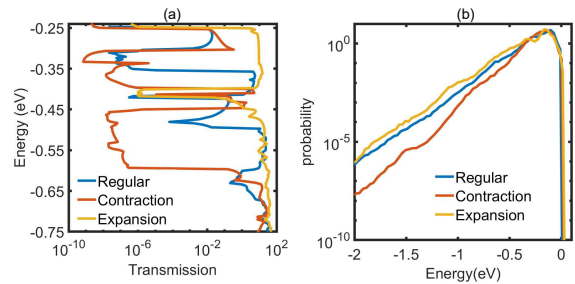}
\caption{\label{carrier_transport} (a) Transmission vs. Energy plot and (b) Carrier Occupation Probability vs. 
Energy for 6ML InAlAs with regular, compressive and tensile strain.}
\end{figure}

\begin{figure*}[t]
\includegraphics[width=0.92\textwidth]{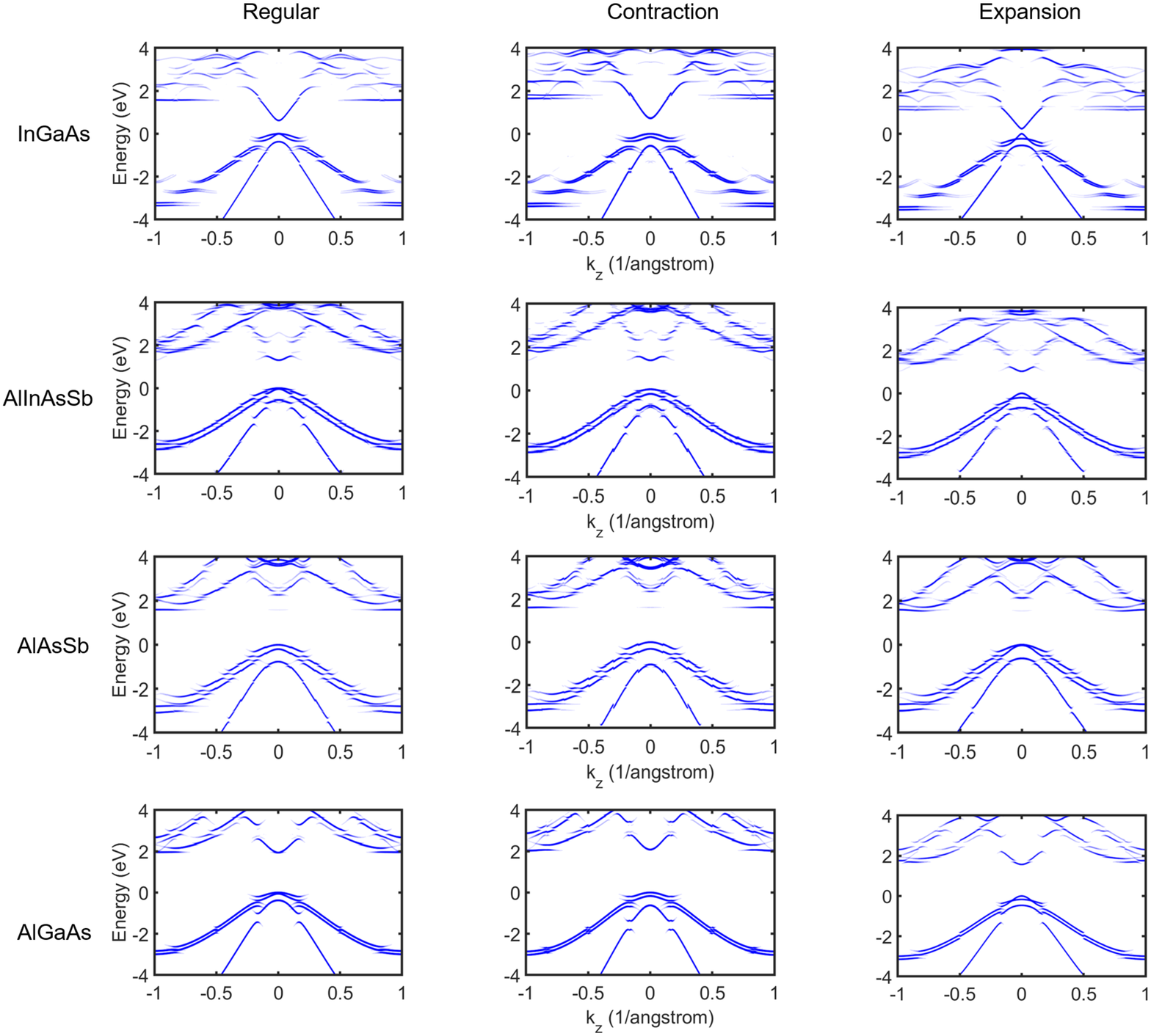}
\caption{\label{allmat_biaxial_stress} Bandstructure of InGaAs, AlInAsSb, AlAsSb and AlGaAs for regular, contraction and expansion cases.
}
\end{figure*}

Having a detailed understanding of the underlying physics of the minigaps, it is then essential to see how can we control the minigap size in these digital alloys. Since, the minigap formation is related to biaxial strain we must study how the bandstructure of these alloys change with strain. We compute the bandstructure for two cases: "contraction"- the substrate lattice constant is smaller than real substrate (InP for InAlAs) lattice constant, and "expansion"- in which the substrate lattice is greater. The bandstructures for strained InAs and  AlAs under contraction and expansion are shown in Fig.~\ref{InAlAs_biaxial_stress}(a) and (b). Under these conditions, the binary constituents experience unequal biaxial strains and due to their different values of Poisson's ratio, $D_{001}$, they also undergo different amounts of strain in the $z$ direction. Thus, their valence bands move by different amounts. InAs has a higher $D_{001}$ than AlAs \cite{Walle1989} and hence bands of InAs are more responsive to strain. We notice that the valence band offset under contraction is large compared to expansion case.   Consequently, the valence bands of InAlAs under contraction become flatter and the minigaps increase in size, as depicted in Fig.~\ref{InAlAs_biaxial_stress}(c). However, we can see in Fig.~\ref{InAlAs_biaxial_stress}(d) that under expansion the InAlAs top valence band effective mass decreases and the  minigaps become smaller. The 2D energy contours of top band of InAlAs in the $x-y$ plane for regular, contraction and expansion cases are depicted in Fig.~\ref{contour_xy}. We observe that for the regular and contraction cases, the top bands are highly anisotropic. Under biaxial strain, in the in-plane ($x$ and $y$) directions the bands move in the opposite direction to that of the out-of-plane ($z$) direction \cite{Nishida2007}. As a result, under contraction the effective mass in the $x-y$ plane decreases. If Fig.~\ref{contour_xy}(b) is compared to Fig.~\ref{contour_xy}(a), we observe the contour lines become more elliptical which indicates the lowering of the mass under contraction. The effective mass increases for expansion as the contour lines becomes flatter in Fig.~\ref{contour_xy}(c). This observed anisotropic nature of the bands can be utilized to explore the use of digital alloys like InAlAs in other applications such as transistors.

One key aspect we need to study is the impact of the strain on carrier transport of digital alloys. Since we have been primarily concerned with the effect of strain in the valence bands, we look at the effect on carrier transport in InAlAs valence band in Fig.~\ref{carrier_transport}. Fig.~\ref{carrier_transport}(a) depicts the ballistic transmssion vs. energy spectrum in the valence band under regular, expansion and contraction conditions in 6ML InAlAs. The transmission has been computed using the NEGF formalism. We observe that as we go from expansion to regular to contraction case the transmission gaps increase in size due to increasing size of the minigaps and enhanced effective mass. As a result, the probability to tunnel across the minigaps decreases and the holes will be more localized near the valence band. This will help in reducing the excess noise in APDs. Another mechanism by which holes can bypass the minigaps is optical phonon scattering. We look at the effect of this scattering using the Boltzmann Transport Equation. The carrier probability vs. energy with optical phonon scattering under an electric field of $1MV/cm$ is shown in Fig.~\ref{carrier_transport}(b). Under expansion condition, holes have a higher probability to occupy higher energy states compared to the regular and contraction cases. Under contraction, the probability is the lowest. Therefore, this is a further indication that contraction prevents holes from reaching higher energies. It is then possible to design better low noise electron injected digital alloy APDs with lower hole impact ionization by applying contraction to materials like InAlAs.

In addition to InAlAs, we also computed the bandstructures of 6ML InGaAs, 10ML AlInAsSb, 5ML AlAsSb and 6ML AlAsSb digital alloys under regular, contraction and expansion conditions along the $001$ direction. These bandstructures are shown in Fig.~\ref{allmat_biaxial_stress}. The primary binary constituents for these alloys are: InAs and GaAs for InGaAs, InAs and AlSb for AlInAsSb, AlAs and AlSb for AlAsSb, and AlAs And GaAs for AlGaAs. For the regular bandstructures, InGaAs and AlAsSb has InP substrate, AlInAsSb has GaSb substrate and AlGaAs has GaAs substrate. The lattice constants for all these materials are taken from the paper by Tan \textit{et al.} \cite{TanETB}. For InGaAs, AlInAsSb and AlAsSb one of the binary constituents has a lattice constant that is greater than the substrate lattice constant while the other constituent lattice constant is smaller. Thus, the binary components experience alternating types of strain. This is not the case for AlGaAs. For all the material combinations, we observe that the effective mass of the top valence band increases under contraction and reduces for expansion. This is mainly because under contraction HH states move up in energy whereas under expansion they move down leaving LH states as the top states in the valence band. For InGaAs, we see that the minigap increases in size with contraction from $0.03 eV$ to $0.16 eV$ which is similar to the behavior of InAlAs described earlier. In AlInAsSb there is a separation between the HH and LH bands around the $\Gamma$ point under contraction. A similar gap is seen for AlAsSb under regular condition. This gap size increases under contraction. The gaps vanish for both AlInAsSb and AlAsSb under expansion. The minigap sizes also increase under contraction by about $0.04 eV$ for AlInAsSb and $0.02 eV$ for AlAsSb. However, for AlGaAs we do not observe any minigaps in the light-hole band. This is primarily because the HH/LH bands of the binary constituents in AlGaAs move in the same direction under strain as both experience same type of biaxial strain. Thus, by band engineering in the digital alloys using biaxial strain, their performance in APDs can be enhanced or possibly used for other applications.

\section{Conclusion}
In this study, we demonstrate that large band offsets result in the formation of minigaps in III-V digital alloys. This band offset results from biaxial strain. Using a orbital chemistry picture, we explained how these minigaps are created. Furthermore, we illustrated that we can engineer the bandstructure by tuning the biaxial strain in wide range of digital alloys. As a general rule, we observe that decreasing the substrate lattice constant can enhance the performance of digital alloys in APDs. 

\section*{Acknowledgment}
This work was funded by National Science Foundation grant NSF 1936016. The authors thank Dr. John P David of University of Sheffield and Dr. Seth R. Bank of University of Texas-Austin for important discussions and insights. The calculations are done
using the computational resources from High-Performance
Computing systems at the University of Virginia (Rivanna)
and the Extreme Science and Engineering Discovery Environment (XSEDE), which is supported by National Science Foundation grant number ACI-1548562.

\providecommand{\noopsort}[1]{}\providecommand{\singleletter}[1]{#1}%

\end{document}